# Spectral signature of periodic modulation and sliding of pseudogap state in moiré system

Yingzhuo Han[1,2,#], Yingbo Wang[1,2,#], Yucheng Xue[1,#], Jiefei Shi[2], Xiaomeng Wang[2], Kenji Watanabe[3], Takashi Taniguchi[4], Jian Kang[5], Yuhang Jiang[2*], Jinhai Mao[1*]

1. School of Physical Sciences, University of Chinese Academy of Sciences, Beijing 100049, China.
2. College of Materials Science and Optoelectronic Technology, Center of Materials Science and Optoelectronics Engineering, University of Chinese Academy of Sciences, Beijing 100049, China.
3. Research Center for Electronic and Optical Materials, National Institute for Materials Science, Tsukuba 305-0044, Japan.
4. International Center for Materials Nanoarchitectonics, National Institute for Materials Science, Tsukuba 305-0044, Japan.
5. School of Physical Science and Technology, ShanghaiTech University, Shanghai 200031, China.

**Abstract:**
The nature of the pseudogap state is widely believed as a key to understanding the pairing mechanism underlying unconventional superconductivity. Over the past two decades, significant efforts have been devoted to searching for spontaneous symmetry breaking or potential order parameters associated with these pseudogap states, aiming to better characterize their properties[1–7]. Recently, pseudogap states have also been realized in moiré systems with extensive gate-tunability, yet their local electronic structure remains largely unexplored[8]. In this study, we report the observation of gate-tunable spontaneous symmetry breaking and sliding behavior of the pseudogap state in magic-angle twisted bilayer graphene (MAtBG) using spectroscopic imaging scanning tunneling microscopy. Our spectroscopy reveals a distinct pseudogap at 4.4 K within the doping range $-3 < \nu < -2$. Spectroscopic imaging highlights a gap size modulation at moiré scale that is sensitive to the filling, indicative of a wave-like fluctuating pseudogap feature. Specifically, the positions of gap size minima (GSM) coincide with regions of the highest local density of states (LDOS) at the filling $\nu = -2.63$, but a unidirectional sliding behavior of GSM is observed for other fillings. In addition, the pseudogap size distribution at certain doping levels also causes a clear nematic order, or an anisotropic gap distribution. Our results have shed light on the complex nature of this pseudogap state, revealing critical insights into the phase diagram of correlated electron systems.

**Introduction:**

Since the discovery of high-temperature cuprate superconductors, the pseudogap and its coexisting intertwined orders have been intensively studied by various experimental techniques[5–7,9–11]. Elucidating the nature of these phases and their relationship with the pseudogap is critical for understanding the pairing mechanism underlying unconventional superconductivity, such as whether the pseudogap represents a precursor pairing state or a competing phase. Recently, the extraordinary superconducting phenomena have also been observed in moiré systems (such as magic-angle twisted bilayer graphene (MAtBG)[8,12–15], rhombohedral trilayer graphene[16], and twisted bilayer tungsten diselenide[17,18]) and attracted significant attention. Similar to cuprate superconductors, the strong correlation of electrons within the flat bands[19–24] is believed to play a crucial role in forming the superconducting state, complicating the understanding of both the origin of superconductivity and the underlying pairing mechanism. Furthermore, the pseudogap state has also been observed in MAtBG at temperatures above the superconducting transition temperature or under an applied magnetic field[8,25–27]. Nano-scale studies, particularly through local probe measurements, may offer valuable experimental insights into the spatial distribution of the pseudogap state and the presence of potential symmetry-breaking phases.

Here we use a spectra-imaging scanning tunneling microscope (SI-STM) to investigate the pseudogap feature in MAtBG and its evolution in real space at different doping levels. We found that the pseudogap state in MAtBG appears above the superconducting transition temperature. Although it occupies the same filling range as superconductivity, the pseudogap nature can be distinguished using point contact spectroscopy (PCS)[8,28]. The pseudogap size is found to depend on two degrees of freedom: real-space position and doping level. Through extensive d$I$/d$V$ mapping at various doping levels, we observe modulations in pseudogap size comparable to the moiré superlattice scales and a peculiar phase sliding phenomenon. At one specific doping level ($v$ = -2.63), the maximum (minimum) gap values are located at positions with the lowest (highest) local density of states (LDOS), leading to a $\pi$ phase difference between the pseudogap size and LDOS modulation. At other doping levels away from

$v$ = -2.63, the maximum (minimum) gap values shift away, indicating a pronounced sliding behavior. Additionally, a significant nematic order in the map of pseudogap size is observed, whereas the map of d$I$/d$V$ intensity retains $C_6$ rotational symmetry. These experimental observations reveal the complex real-space distribution of pseudogap states at different doping levels, offering deeper insights into the nature of unconventional superconductivity in moiré systems.

**Results:**

**Distinguishing the pseudogap feature**

Fig. 1a illustrates the device setup utilized in this study to investigate MAtBG through the STM, where the tip is grounded and bias voltage is applied to the sample. The MAtBG sample is supported by a hexagonal boron nitride (hBN) substrate and placed on a Si/SiO$_2$ wafer, allowing carrier density tuning via gate voltage $V_g$ applied to the Si substrate. The STM topography in Fig. 1b reveals the typical moiré pattern of MAtBG, with the black dashed line delineating one moiré unit cell. The red, blue, and yellow circles denote three distinct stacking regions: AA, AB, and BA, respectively. According to the measured moiré length $l_M$ = 13.44 nm, a twist angle of 1.05° can be extracted, placing our sample within the 'magic angle' regime[29]. In this regime, the Fermi velocity of the electrons almost vanishes, leading to two weakly dispersive flat bands (as shown in the calculated band structure in Fig. 1c), where electron-electron interaction plays a vital role in the electronic ground state. Moreover, each flat band has fourfold degeneracy, corresponding to isospin (spin and valley) degrees of freedom[30–32].

We utilize scanning tunneling spectroscopy (STS) to investigate the evolution of these flat bands under different doping levels. Firstly, as shown in Fig. 1e, each flat band produces a peak in the d$I$/d$V$ spectrum[21–23,33]. At the charge neutral point (CNP), two peaks are symmetrically positioned on either side of the Fermi level ($V_{Bias}$ = $E_F$ ≡ 0 meV). When the flat bands become fully occupied (empty), the respective peaks shift to the left (right) of the $E_F$, respectively. The substantial difference in the spacing between the two flat bands at empty state ($\Delta E_{FB}$ = 23 meV) and at CNP ($\Delta E_{FB}$ = 49 meV) provides compelling evidence of strong electron correlations[20,34]. The filling factor is defined as $v = -4$ when two flat bands are both empty (green curve) and $v = +4$ when two flat bands are fully occupied (red curve). Following this criterion, we could determine all the intermediate filling factors $v$ precisely in Fig. 1d.

By continuously tuning the carrier density via $V_g$ and measuring d$I$/d$V$ spectroscopy, we observe two flat bands sequentially pass through the $E_F$ (Fig. 1d). At each integer filling, consistent with previous studies, the cascade filling feature of the flat bands appears (indicated by the black dashed line), attributed to the lifting of isospin degeneracy due to strong electron-electron interactions[35–37]. The most striking feature emerges within the doping range of $-3 < v < -2$, where a clear gap feature is observed at $E_F$ in both AA and AB stacking regions (green arrows). Previous research has demonstrated the emergence of unconventional superconductivity in this doping range[8,12]. Similar gap features have been visualized in MAtBG sample when superconductivity is suppressed either at temperatures above the transition temperature ($T_c \sim 1.7$ K) or in the presence of a magnetic field[8]. Given that our base temperature is 4.4 K, we attribute this gap to a pseudogap, which is further confirmed by the PCS and temperature-dependent measurements.

To provide a more detailed examination of this pseudogap feature, Fig. 1f shows the zoomed-in gate-dependent d$I$/d$V$ spectra for both AA and AB stacking regions with finer gate voltage intervals and higher energy resolution. The pseudogap is manifested as a suppression of d$I$/d$V$ intensity at $E_F$, further supporting its correlation-induced nature. Fig. 1g presents the d$I$/d$V$ line cuts as a function of gate voltage ranging from $V_g = -12$ V to 26 V (left) and the corresponding second derivative of -d$I$/d$V$ curves (right). This data processing method effectively mitigates the impact of the background intensity, particularly from the nearby flat bands, while enhancing the visibility of the pseudogap features without altering intrinsic gap size[38,39] (see Supplementary Information (SI), Section II for more comparison). The evolution of peaks flanking the pseudogap with the gate voltage can be clearly identified in the right panel (highlighted by the black dashed lines). We further compare the pseudogap features in the AA and AB regions and observe distinct evolution processes in each area. Fig. 1h shows the single d$I$/d$V$ spectrum at one typical gate voltage in Fig. 1f ($V_g$ = -17 V, corresponding to $v$ = -2.45), which exhibits a clear gap-size mismatch. This difference suggests a modulation of the pseudogap size within the moiré scale, which may vary, leading to distinct modulation patterns under different doping conditions (see SI, Section II).

Although both states exhibit gap features, the pseudogap state fundamentally differs from the superconducting state due to its distinct physical mechanism. Consequently, point contact spectroscopy (PCS) measurements in STM experiments are commonly employed to distinguish them[8,40]. In PCS measurements of the SC state, as the tip approaches or even touches the SC surface, electrons from the metallic tip can be reflected as holes, with Cooper pairs propagating into the superconducting sample. This process, known as Andreev reflection, results in an intensity enhancement in the d$I$/d$V$ spectrum near $E_F$ [41,42]. However, in the pseudogap state, the absence of coherent Cooper pairs precludes Andreev reflection, leaving the gap features unchanged in the d$I$/d$V$ spectrum near $E_F$.

We performed PCS measurements on the same sample at both 4.4 K and 0.4 K to compare these two distinct states in Fig. 2. We first cool down the sample to 0.4 K to identify the superconducting gap state. As we gradually increase the tunneling current setpoint to bring the tip closer to the sample surface, the two coherence peaks gradually come close together and eventually merge into one single peak at $E_F$ (Fig. 2a), which is indicative of Andreev reflection. From the individual d$I$/d$V$ spectra curves obtained with progressively higher setpoint values (Fig. 2b), a clear transition from the normal tunneling regime to the point contact regime can be observed, consistent with the schematic illustration in Fig. 2c. When we apply an increasing out-of-plane magnetic field $B$, the Andreev reflection peak is gradually suppressed and finally turns into a pseudogap feature beyond the critical value $B_c$ ~ 50 mT (Fig. 2d). Previous transport experiments[12] measured this value as tens of mT, which is of the same order of magnitude as our experimental results. Fig. 2e employs Dynes-function to fit the gap in the tunneling spectrum measured at $v$ = -2.6 using the model quasiparticle DOS for both an *s*-wave nodeless superconductor (grey curve) and a nodal superconductor (red curve). The fitting yields a superconducting order parameter of 2$\Delta$' = 2.20 meV for *s*-wave superconductor and 2$\Delta$' = 3.06 meV for nodal superconductor (see SI, Section III for fitting details). The STS spectra align more closely to a nodal superconductor, such as higher-angular-momentum pairing (e.g., *p*- or *d*-wave) characterized by an anisotropic gap function. This nodal superconducting gap characteristics have also been validated in graphene-based moiré systems through transport experiments[43,44]. All the behavior mentioned above is consistent with the previous transport[12,13,15] and probe[8] studies, providing strong evidence for the superconducting state at 0.4 K. However, the PCS measurements conducted at 4.4 K yield distinctly different results. As the tip gets close to the sample, the gap feature at $v$ = -2.6 persists even at a tunneling current of 150 nA (Fig. 2f and 2g), supporting our hypothesis that this gap corresponds to a pseudogap state. Considering the contrasting results at the two temperature intervals, we can confidently attribute the gap feature detected at 4.4 K to the pseudogap state.

Additionally, we measured the temperature-dependent d$I$/d$V$ spectrum across the 4.6 K to 9.1 K range (Fig. 2h). As the temperature rises, the two peaks of pseudogap gradually converge, with the gap feature ultimately disappearing at approximately 7 K. Furthermore, gate-dependent d$I$/d$V$ spectrum at 9 K confirms the lack of pseudogap feature within the doping range of -3 < $v$ < -2, while the cascade features remain unaffected (Fig. 2i and 2j). The absence of both pseudogap and superconducting gap at higher temperatures further supports their correlation-induced nature, underscoring their similarity to those observed in cuprate superconductors[45].

**Imaging spatial gap size modulation**

Now we focus on the spatial variation of the pseudogap size (2$\Delta$). We conducted real-space d$I$/d$V$ mapping at five different doping levels and extracted the distribution of the pseudogap size ($\Delta$-maps). Due to space limitations, results for four doping levels are presented in Fig. 3b-e, with corresponding topography shown in Fig. 3a; the data for $v$ = -2.40 are provided in SI, Section VI. The black dashed diamonds in each graph delineate the same moiré unit cell within the STM topography, with vertices corresponding to the AA stacking regions. The pseudogap sizes exhibit apparent inhomogeneity at all five doping levels, forming a triangular lattice at the moiré superlattice scale. In this arrangement, purple rounded regions correspond to gap size minima (GSM), while surrounding orange regions represents larger gap sizes. As mentioned earlier, the AA stacking region is the highest point in terms of topography and is also the area where the LDOS of the flat band is primarily concentrated, whereas the LDOS of the flat band in the AB region is significantly reduced. Given this substantial modulation in the LDOS, the emergence of such real-space modulation is possible[6]. Nevertheless, the GSM does not consistently align with the LDOS configuration across five different fillings. Especially at $v = -2.63$, the GSM is located precisely on the AA stacking region, resulting in a significant $\pi$ phase shift between LDOS and pseudogap size (can be confirmed by the GSM precisely aligning with the vertices of the black rhombus in Fig. 3c). With this doping level as the reference, in the relatively lower hole-doping ($v = -2.45$, in Fig. 3b), the GSM shifts upward to the left along one highly symmetric direction relative to the AA stacking region; while in the relatively higher hole-doping ($v = -2.81$ and $v = -2.90$, in Fig. 3d and 3e), the GSM shifts in precisely the opposite direction (i.e. down-right). This sliding behavior highlights the significant dependence of the spatial distribution of the pseudogap on varying doping levels.

To further quantify this sliding behavior, we extract the pseudogap sizes ($2\Delta$) along the highly symmetric direction where GSM sliding occurs (indicated by the green arrows in each map), and compare them with the corresponding topographic heights ($Z$), which also reflect the modulation of LDOS. As shown in Fig. 3f-i, both $2\Delta$ and $Z$ exhibit sinusoidal modulation with a periodicity comparable to that of the moiré lattice. Notably, a phase difference is observed between these two quantities. We define $\delta_d$ as the difference between the position of the GSM and the center of the AA stacking region, and $\delta_{phase}$ as $2\pi \times \frac{\delta_d}{L_{moire}}$ to represent the relative phase shift due to the spatial sliding. The $\delta_{phase}$ values as a function of $v$ are plotted in Fig. 3k. At $v = -2.63$, the $\delta_{phase}$ is almost zero, indicating a good alignment of the GSM with the AA stacking region. While at $v = -2.40$ and $v = -2.45$, the $\delta_{phase}$ approach $\pi/2$, indicates a noticeable sliding of the GSM away from the AA region in a specific direction. Conversely, at $v = -2.81$ and $v = $

-2.90, the $\delta_{\text{phase}}$ is approximately $-\pi/2$, signifying sliding of GSM in the opposite direction. These results highlight the doping-dependent nature of the spatial distribution of the pseudogap size. Based on prior research on superconductivity in MAtBG[8,13], we find that $v$ = -2.63 is near the "optimal" doped regime, where "optimal" means the superconducting transition temperature is the highest. While $v$ = -2.40 ($v$ = -2.90) correspond to the under-doped (over-doped) regimes, respectively. Our results highlight that the pseudogap modulation is not locked to the underlying moiré superlattice, but responses sensitively to the filling of the flat bands, presenting the wave-like fluctuating behavior.

**Rotational symmetry-breaking in $\Delta$-maps**

Besides the $\delta_{\text{phase}}$ oscillation described above, signatures of rotational symmetry-breaking have also been revealed in two distinct ways. Firstly, the GSM sliding occurs in a unidirectional manner. For instance, at the filling factor of $v$ = -2.45, the GSM is situated away from the AA stacking region along a specific direction. Upon rotating the underlying moiré superlattice by 60°, while the AA regions are well repeated, the positions of the GSM do not coincide, thereby indicating the breaking of the $C_6$ rotational symmetry. Another way of rotational symmetry-breaking can be directly observed from the real-space distribution of the pseudogap at some doping levels. For example, in Fig. 3d and 3e, the pseudogap exhibits a distinct nematic order, with the orientations differing. To further elucidate this characteristic, we applied the self-correlation analysis to the $\Delta$-maps (details see SI, Section VII). This analytical approach yields the correlation coefficients of the image with itself, and has therefore been used to amplify subtle symmetry and periodicity features that may be otherwise difficult to detect[46]. Fig. 4a and 4b show the $\Delta$-map taken at $v$ = -2.63 ("optimal" doped regime) and the corresponding self-correlation image. Each GSM region exhibits an almost circular shape with preserved $C_6$ rotational symmetry, as clearly reflected by the separated bright circles in Fig. 4b. However, $\Delta$-map taken at $v$ = -2.81 (Fig. 4e) shows contrasting results, where the purple GSM regions take on an ellipse shape, with the interspersed orange areas displaying a nematic distribution. These features are manifested as one-dimensional bright lines in the self-correlation image (Fig. 4f), indicating a disruption of $C_6$ rotational symmetry (For the same analysis at more doping levels, see SI, Section VIIII).

A crucial question is whether the rotational symmetry-breaking observed here has intrinsic or extrinsic origins. To address this, we extracted the d$I$/d$V$ maps at the energy within the pseudogap ($E$ = -1 meV), and performed the self-correlation analysis (as illustrated in Fig. 4c, d for $v$ = -2.63 and 4g, h for $v$ = -2.81). Notably, no signature of nematic order was detected in this LDOS channel in either case, suggesting that symmetry breaking is not a global external effect that leads to Fermi surface distortion. This makes the spontaneous symmetry-breaking phase within the pseudogap a more plausible explanation. Moreover, as shown in Fig. 4i, we identified a transition region where the nematic direction undergoes a sharp switch. Although the STM topography of this region (inset in Fig. 4i) reveals a uniformly $C_6$ symmetric moiré pattern, the two areas separated by the red dashed line exhibit distinct nematic axes, as indicated by the orange and green dashed arrows. Furthermore, we divided the image along the red dashed line and independently performed self-correlation analysis on each section, with the results aligning well with our hypothesis (Fig. 4j and 4k). The presence of such a domain wall not only reinforces the validity of spontaneous symmetry breaking but also suggests that the emergence of the nematic axis is not given by any global factors. Therefore, we can essentially rule out the extrinsic origins (such as strain, a common factor causing symmetry-breaking in two-dimensional materials), and instead focus on other intrinsic physical characteristics related to the origin of the pseudogap.

**Discussion**

Currently, a key challenge in understanding the pseudogap in cuprates is clarifying its relationship with coexisting symmetry-breaking states, such as charge density waves (CDW)[47] or spin density waves (SDW)[48], which have been extensively observed through STM and Angle-Resolved Photoemission Spectroscopy (ARPES) measurements. Some researchers propose the pseudogap arises from unidirectional CDW as a competing phase to superconductivity[7], while others question if the CDW state is sufficiently strong to fully account for the pseudogap[10,49]. In the MAtBG system, there also exits the moiré-induced LDOS modulation and the nematic order that embedded in the pseudogap size distribution; however, significant differences remain compared to those observed in cuprate superconductors. Firstly, in cuprates, the pseudogap regions appear at the lowest doping levels as nanometer-scale clusters, indicating their short-range localized nature[6,38]. In contrast, the pseudogap in MAtBG exhibits a nearly global characteristic with a typical size extending up to hundreds of nanometers, which is even larger than the superconducting coherence lengths measured by transport measurements. Consequently, the scenario of electron pre-pairing above the critical temperature of superconductivity appears less plausible as an explanation. Additionally, our PCS measurements demonstrate that the pseudogap features persist at higher magnetic fields, further ruling out the pre-paring mechanism. Secondly, the ordered state within the pseudogap phase in cuprates breaks translational symmetry, predominantly manifesting as a 'checkerboard' pattern with a periodicity of $4a_0$ (ref[1-3,6,7,38]). However, in the pseudogap phase of MAtBG, no significant translation breaking is observed for the charge density on the moiré scale, while nematic sliding observed between the gap size distribution and the LDOS hints at redistribution of the charge density within each moiré unit cell at different doping levels, or a charge bond order. Third, the nematicity of the pseudogap phase is ubiquitous in the under-doped regime of cuprate superconductors[5-7,38,50], while in MAtBG, nematic order is absent at certain doping levels (e.g. $v$ = -2.63). Moreover, nematic ordered regions emerge as clusters embedded within $C_4$-symmetric matrices of a weak insulating phase in cuprates, with spatial variations of the LDOS exhibiting similar structural patterns[6]. In MAtBG, the nematic order emerges in pseudogap size maps that is uniform at the hundreds of nanometers scale (see SI, Section VIII), while the LDOS distribution retains $C_6$ rotational symmetry, showing no signs of similar nematicity.

These differences highlight that while MAtBG shares some analogies with cuprate superconductors, it also possesses distinctive characteristics. Notably, even though the wave-like pseudogap modulation is observed in all the doping range, the nematicity is absent near the optimal doping regime ($v \sim$ -2.63), which indicates that the symmetry breaking in the pseudogap regime may tune the emergence of the superconductivity at lower temperature.

The emergence of the nematicity does not cause an apparent translational symmetry breaking, therefore theoretically, it does not necessarily open a gap and thus cause the metal to insulator transition. It is plausible that the measured gap in the pseudogap phase results from another long-range order in the charge, spin, or valley channels. Although this order cannot be directly observed in the STM measurements, it is supported by the periodically modulated rather than uniformly distributed pseudogap observed in the real space. Moreover, the gap measured in the SC state is much larger than $k_B T_c$ ($T_c \sim 2$ K) and has roughly the same magnitude as the one in the non-SC state, suggesting that the gaps found in non-SC and SC states result from the same long-range order. Therefore, two orders, one rooted from the non-SC pseudogap state and much stronger than the pairing one, coexist in the SC states, while the former non-SC order dominates in the STM measured gap. Whether these two orders compete or cooperate still remain unknown. Our work has led to the realization that the pseudogap phase could be associated with a variety of competing orders, which may coexist with superconductivity and can influence its emergence. It also broadens the implications of pseudogap physics and its potential unifying role across different material systems.

**Method:**

**Sample fabrication.** The devices were fabricated using a "tear-and-stack" method[51]. Monolayer graphene (MLG) and few-layer hexagonal boron nitride (hBN) were first exfoliated onto $SiO_2$/Si substrates and characterized using optical and atomic force microscopes. A polypropylene carbonate (PPC) film and polydimethylsiloxane (PDMS) were used to sequentially pick up the exfoliated flakes—hBN, half of the MLG, and then the remaining half of the MLG. Before picking up the second half of the MLG, the stage was manually rotated by approximately 1° to achieve the desired twisted bilayer graphene (tBG) structure. All pick-up procedures were performed at temperatures between 40-50°C. After assembling the heterostructure, the PPC film was peeled off, flipped, and transferred onto a $SiO_2$/Si substrate ($SiO_2$ thickness: 285 nm) at 105°C. The PPC film was then removed by high-vacuum annealing at 350°C. To facilitate scanning tunneling microscopy (STM) measurements, the heterostructures were patterned into gate-tunable devices using e-beam lithography followed by Au/Ti metal evaporation. The devices were then annealed in an ultra-high vacuum chamber at 400°C overnight to remove any residual before the STM measurements.

**STM measurements.** The STM experiments are performed at 4.4 K with an ultrahigh vacuum better than $1 \times 10^{-10}$ mbar. The STM tips are calibrated on Cu(111) before use. The topographies are obtained in constant current mode. The d$I$/d$V$ spectra and maps are collected by standard lock-in technique, with 0.5 mV AC modulation at 711 Hz added to the DC sample bias. The low temperature data were acquired at 0.4 K by $^3$He evaporation. The magnetic field is provided by the current in the superconducting coil, and the direction of the magnetic field is perpendicular to the sample surface.

**Data Availability:**
Relevant data supporting the key findings of this study are available within the article and the Supplementary Information file. All raw data generated during the current study are available from the corresponding authors upon request.

**Acknowledgements:**
We appreciate the enlightening discussions with Prof. Kun Jiang, Prof. Fuchun Zhang, and Prof. Jianxin Li.

**Author contributions:**
J.M. and Y.J. conceived and supervised the experiments. Y.H., J.S., and X.W. were responsible for device fabrication. Y.H. and Y.W. conducted the STM measurements, assisted by Y.X. K.W. and T.T. provided the hBN crystals. All authors participated in discussions of the results and contributed to the manuscript preparation.

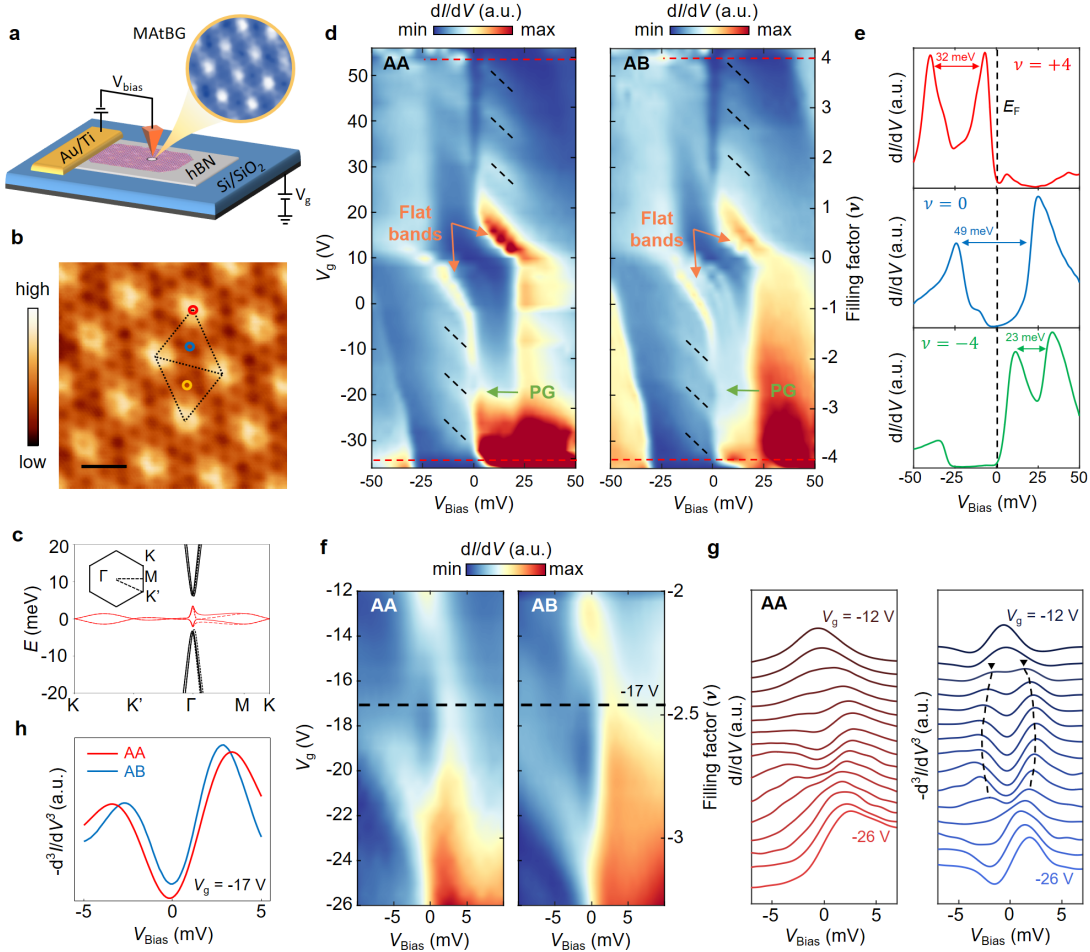

**Fig. 1 Electronic structure of MAtBG and pseudogap states at different stacking regions. a,** The schematic of the device and STM experimental setup. $V_{Bias}$ is the bias voltage applied between the sample and tip, and $V_g$ is the gate voltage used to tune the carrier concentration. **b,** STM topography of the MAtBG ($V_{Bias}$ = -100 mV, $I$ = 0.1 nA; scale bar, 10 nm). The brightest region is the AA stacked region (red), the darkest triangular region is the AB/BA stacked region (blue / yellow), and the domain wall (dashed line) structure connects the AA stacked region. **c,** Band structure calculated using a continuum model for MAtBG ($\theta$ = 1.05°), with red color highlighting two flat bands. Solid and dashed lines correspond to the band of **K** and −**K** valley. The inset provides a sketch of its Brillouin zone. **d,** Gate-dependent d$I$/d$V$ spectra of the AA (left) and AB (right) stacked region ($V_{Bias}$ = -50 mV, $I$ = 0.4 nA). The flat bands / pseudogap (PG) are indicated by orange / green arrows. The black dashed lines label the cascade features and the red dashed lines indicate the filling factors as ±4. **e,** d$I$/d$V$ spectra taken in the AA regions for the fully occupied band (red curve, $\nu$ = +4), charge neutral point (blue curve, $\nu$ = 0), and fully empty band (green curve, $\nu$ = -4). **f,** High-resolution d$I$/d$V$ spectra for the AA and AB stacked regions at $V_g$ ranging from -12 to -26 V. **g,** Left, d$I$/d$V$ line cuts for the AA regions at the same $V_g$ in **f**. Right, -d$^3I$/d$V^3$ curves of the spectra in the left panel. The black dashed lines and arrows indicate the evolution of two peaks. **h,** -d$^3I$/d$V^3$ spectra for the AA (red curve) and AB (blue curve) regions at $V_g$ = -17 V.

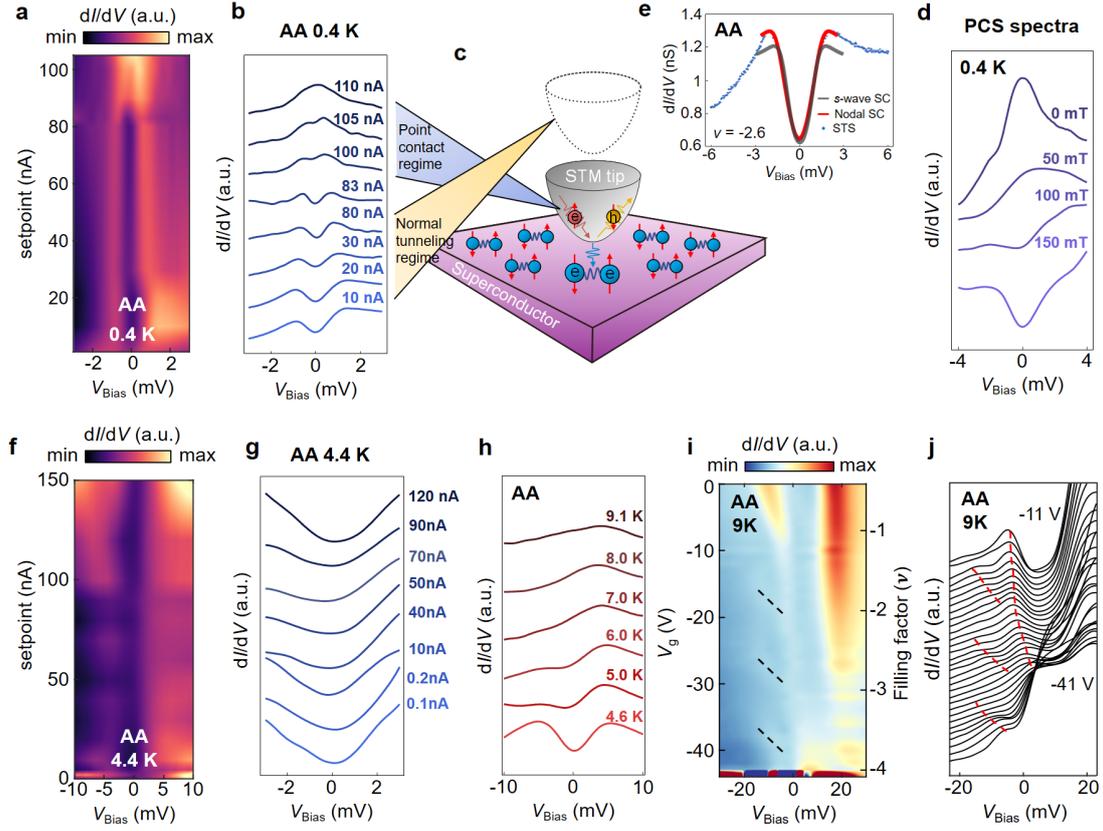

**Fig. 2 PCS measurements to identify gap states at $v = -2.6$. a,** Point contact spectra (PCS) at the superconducting phase ($v = -2.6$) for the AA stacked region at 0.4 K. The spectra are normalized for clarity. **b,** Normalized setpoint-dependent $dI/dV$ line cuts ranging from $I = 10$ nA to 110 nA. When setpoint exceeds 80 nA, the tip transitions from the normal tunneling regime to the point contact regime, with an intensity enhancement occurs near $E_F$. **c,** A schematic diagram illustrates Andreev reflection occurring at the interface between a metallic STM tip and an SC sample with two regimes. **d,** Magnetic-field dependent PCS $dI/dV$ spectra at the superconducting phase ($v = -2.6$). Curves are normalized for clarity. **e,** Dynes-function fits to the experimental tunneling spectrum (blue dots) measured at $v = -2.6$ at AA stacked region at 0.4 K using the model quasiparticle DOS for a nodeless *s*-wave superconductor (grey curve) a nodal superconductor (red curve). The STS spectra is clearly more consistent with the nodal superconductor fitting. **f,** PCS at the pseudogap phase ($v = -2.6$) for the AA stacked region at 4.4 K. The spectra are normalized for clarity. **g,** Normalized setpoint-dependent $dI/dV$ line cuts ranging from $I = 0.1$ nA to 120 nA. **h,** Temperature-dependent $dI/dV$ spectra for the pseudogap state ($v = -2.6$) of AA stacked region. **i,** Gate-dependent $dI/dV$ spectra taken at 9 K. The pseudogap state is absent between $v = -2$ and $-3$. Cascade features are indicated by the black dashed lines. **j,** Detailed gate-dependent $dI/dV$ intensity plot for the gate voltage range same as **i**. Red dashed lines labeled the cascade feature (isospin flavors) and the $dI/dV$ intensity near $E_F$.

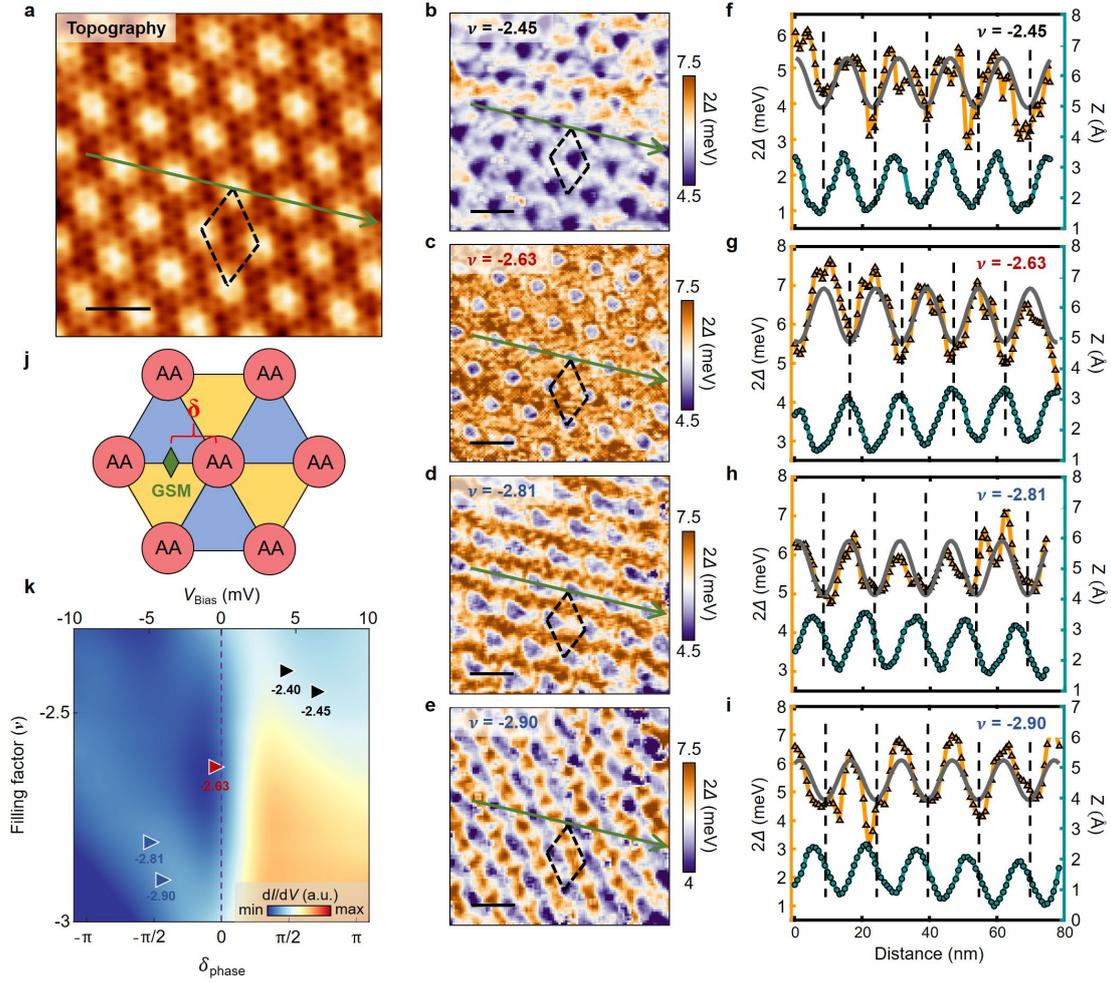

**Fig. 3 Spatial variations of the pseudogap. a,** The STM topography of 80×80 nm (scale bar, 16 nm). The dashed diamond represents a moiré unit cell, which is also used as a marker to label the same position in **b-e**. The green arrow represents the special highly symmetry direction where the gap size minima (GSM) sliding occurs. **b-e,** $\Delta$-maps taken from the same area as shown in **a** at $v$ = -2.45 (**b**), $v$ = -2.63 (**c**), $v$ = -2.81 (**d**), and $v$ = -2.90 (**e**). **f-i,** The extracted gap sizes (orange curve) and the topographic heights (turquoise curve) along the green arrow in **b-e** correspondingly. The grey line is fitted curve using a sine function to trace the oscillating gap. The black dashed line indicates the position of the GSM and its corresponding location on the topographic height. **j,** The schematic diagram of the definition of $\delta_{phase}$, which represents the relative position of GSM compared with the center of AA stacking region (i.e. the highest point in topography) within one moiré unit cell. **k,** The extracted $\delta_{phase}$ at different $v$ (marked as solid triangles). The background corresponds to the d$I$/d$V$ spectra of the AB stacking region at filling factors $v$ = -2.3 ~ -3. It can be observed that the pseudogap is relatively larger at $v$ = -2.63 (the "optimal" filling), while it is smaller at the other four filling factors.

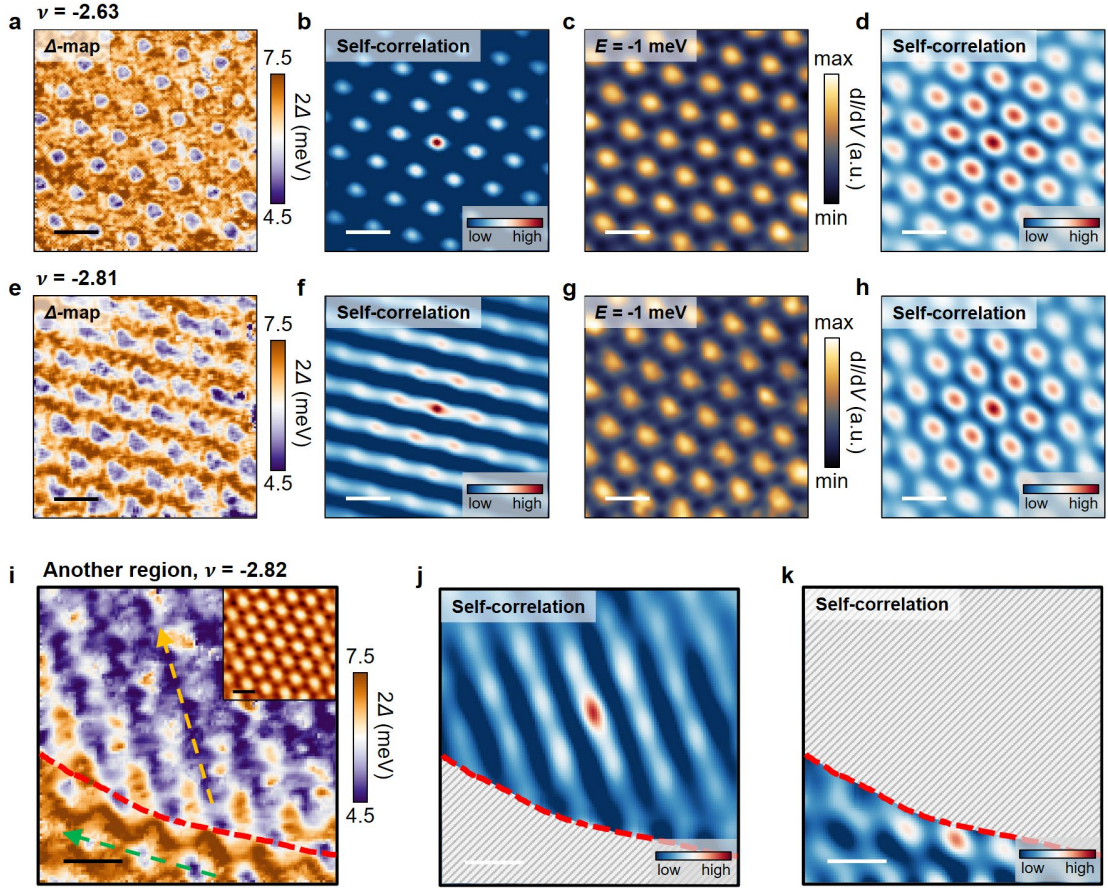

**Fig. 4 Visualization of the nematic order in $\varDelta$-map. a,b,** $\varDelta$-map and the corresponding self-correlation image at $v = -2.63$, exhibiting retained $C_6$ rotational symmetry. **c,d,** d$I$/d$V$ intensity map at energy within the gap ($E = -1$ meV) and the corresponding self-correlation image, taken in the same area and doping level as **a**. **e-h,** Same as **a-d**, but taken at $v = -2.81$. The $\varDelta$-map in **e** exhibiting clear nematic order, which is further revealed in the self-correlation image in **f**. **i-k,** $\varDelta$-map and the corresponding self-correlation image taken at anther region (region B) at $v = -2.82$. The red dashed line separate two areas with distinct nematicity axes (pointed by the orange and green dashed arrows in **i**), consistent with the results shown in the self-correlation image in **j** and **k**. The inset in **i** displays the STM topography of the region B. The scale bars in all the images are 16nm.